\documentclass[conference]{IEEEtran}
\usepackage{lipsum}
\IEEEoverridecommandlockouts

\usepackage{cite}
\usepackage{amsmath,amssymb,amsfonts}
\usepackage{moresize}
\usepackage{float}
\usepackage{graphicx}
\usepackage{textcomp}
\usepackage{xcolor}
\usepackage{url}
\usepackage{braket}
\usepackage{parskip}
\usepackage[switch]{lineno}
\usepackage{algorithm}
\usepackage{algpseudocode}
\usepackage{nicefrac}
\usepackage{siunitx}
\usepackage{tikz}
\DeclareRobustCommand{\circnum}[1]{\tikz[baseline=(char.base)]{%
  \node[shape=circle,fill=black,inner sep=1pt,font=\sffamily\scriptsize\color{white}] (char) {#1};}}
\usepackage{newtxtext} 
\usepackage{newtxmath} 
\usepackage[expansion=true,protrusion=true]{microtype}
\usepackage{etoolbox}
\usepackage{flushend}
\usepackage{needspace}

\usepackage{titlesec}
\newlength{\secskipbefore}
\newlength{\secskipafter}
\titlespacing*{\section}      {0pt}{\secskipbefore}{\secskipafter}
\titlespacing*{\subsection}   {0pt}{\secskipbefore}{\secskipafter}
\titlespacing*{\subsubsection}{0pt}{\secskipbefore}{\secskipafter}

\makeatletter
\def\@IEEENORMtitlevspace{0pt}
\def\@IEEEMINtitlevspace{0pt}
\def\@IEEEauthorblockconfadjspace{0.5em}
\def\@IEEEauthorblockNstyle{\normalfont\fontsize{10.45pt}{12.54pt}\selectfont}
\def\@IEEEauthorblockAstyle{\normalfont\fontsize{10.45pt}{12.54pt}\selectfont}
\patchcmd{\@maketitle}{\addvspace{0.5\baselineskip}}{\vskip-0.5\baselineskip}{}{}
\makeatother
\def\IEEEtitletopspace{0pt}

\def\BibTeX{{\rm B\kern-.05em{\sc i\kern-.025em b}\kern-.08em
    T\kern-.1667em\lower.7ex\hbox{E}\kern-.125emX}}

\begin{document}

\title{{\fontsize{19pt}{22pt}\selectfont\textls[-50]{Robust GHZ State Preparation via Majority-Voted Boundary Measurements}}}
\author{

 \IEEEauthorblockN{Jean-Baptiste Waring}
 \IEEEauthorblockA{\textit{\footnotesize Dept. of Electrical \& Computer Engineering} \\
 \textit{Concordia University}\\
 Montréal, Québec \\
 j\_warin@live.concordia.ca}
 \and
 \IEEEauthorblockN{Sébastien Le Beux}
 \IEEEauthorblockA{\textit{\footnotesize Dept. of Electrical \& Computer Engineering} \\
 \textit{Concordia University}\\
 Montréal, Québec \\
 sebastien.lebeux@concordia.ca}
 \and  
 \IEEEauthorblockN{Christophe Pere}
 \IEEEauthorblockA{\textit{\footnotesize Dép. de génie logiciel et des technologies de l'information} \\
 \textit{INTRIQ, ÉTS Montréal,}\\
 Montréal, Québec \\
 christophe.pere@etsmtl.ca}
}
\maketitle

\begin{abstract}
Preparing high-fidelity Greenberger-Horne-Zeilinger (GHZ) states on noisy quantum hardware remains challenging due to cumulative gate errors and decoherence.
We introduce Group-Majority-Voting (Group-MV), a dynamic-circuit protocol that partitions arbitrary coupling graphs, prepares local GHZ states in parallel, and fuses them via majority-voted mid-circuit measurements.
The majority vote over redundant boundary links mitigates measurement errors that would otherwise propagate through classical feedforward.
We evaluate Group-MV on simulated Heavy-hex and Grid topologies for 30 through 60 qubits under a realistic noise regime.
Group-MV generalizes to arbitrary GHZ sizes on arbitrary coupling topologies, achieving $2.4{\times}$ higher fidelity than the Line Dynamic method while tracking the unitary baseline within 3\%.
\end{abstract}

\vspace*{10pt}\vspace*{-\baselineskip}

\section{Introduction}

Realizing the full potential of quantum computing relies on the ability to generate and preserve high-quality entanglement across many qubits, a capability that underpins virtually all quantum algorithms~\cite{hoyer_quantum_2005,pham_2d_2013}. On current Noisy Intermediate-Scale Quantum (NISQ) devices~\cite{preskill_quantum_2018}, however, entanglement quality degrades as noise and decoherence accumulate through each successive gate layer~\cite{burnett_decoherence_2019}. 
Greenberger-Horne-Zeilinger (GHZ)~\cite{kafatos_going_1989} states are maximally entangled $N$-qubit states and provide a natural benchmark for quantum computer's ability to prepare large-scale entanglement.
These states are foundational resources across quantum information. For example, they enable quantum metrology applications by leveraging the collective phase accumulation of the entangled state~\cite{wei_verifying_2020} and generally serve as benchmarks for characterizing multi-qubit device quality, usually using state fidelity.
Preparing GHZ states on hardware, however, is constrained by \textit{i)} a fundamental $\mathcal{O}(\log n)$ lower bound on two-qubit gate depth under unitary evolution, even with all-to-all connectivity~\cite{baumer_efficient_2024}, \textit{ii)} further depth scaling up to $\mathcal{O}(n)$ for linearly connected hardware topologies, and \textit{iii)} cumulative decoherence and gate errors that compound with each additional layer~\cite{ozaeta_decoherence_2019}. Together, these constraints degrade the prepared state. 
These limitations can, in principle, be overcome by dynamic circuits, which replace part of the quantum gate depth with mid-circuit measurements and classically conditioned feedforward operations. This approach enables constant-depth preparation of $\ket{\textrm{GHZ}_N}$ states \cite{baumer_efficient_2024}. 
Similarly, the network-oriented approach of~\cite{ghaderibaneh_generation_2023}, which partitions and fuses local GHZ states to optimize generation rates over stochastic channels, does not translate directly to monolithic quantum processing units.
No existing method combines the parallelism benefits of partition-and-fuse with the constraints of arbitrary single-device coupling maps. Furthermore, current measurement-based approaches condition feedforward on a single measurement and are thus very sensitive to mid-circuit measurement error propagation.

Intuitively, partitioning offers a natural path to scalability: by preparing smaller GHZ states in parallel and fusing them, we trade sequential depth for a bounded number of mid-circuit measurements.
Each local preparation operates on $K$ qubits with depth $\mathcal{O}(\log K)$, independent of the total system size $N$.
When mid-circuit measurement errors dominate, majority voting provides a straightforward error-tolerance mechanism absent from purely unitary approaches.

In this paper, we introduce \emph{Group-Majority-Voting} (Group-MV), the first protocol for robust GHZ state preparation on arbitrary single-device topologies using a partition-and-fuse strategy. Our contributions are:
\begin{itemize}
    \item Majority-voted boundary measurements for robust classical feedforward, mitigating mid-circuit measurement errors,
    \item an algorithm to automatically partition $N$ qubits into groups of target size $K$ ensuring a minimum boundary size of $L$ between groups,
    \item an algorithm to synthesize coupling-graph-compliant GHZ state preparation Group-MV circuits using Qiskit.
\end{itemize}
Results demonstrate that Group-MV generalizes to arbitrary GHZ sizes on arbitrary coupling topologies. Across Heavy-hex and Grid (square lattice) architectures ranging from 30 to 60 qubits, Group-MV with $L{=}3$ achieves $2.4{\times}$ higher fidelity than the state-of-the-art Line Dynamic method while tracking the unitary baseline within 3\%.

\begin{figure*}[ht!]
    \centering
    \includegraphics[width=0.84\linewidth]{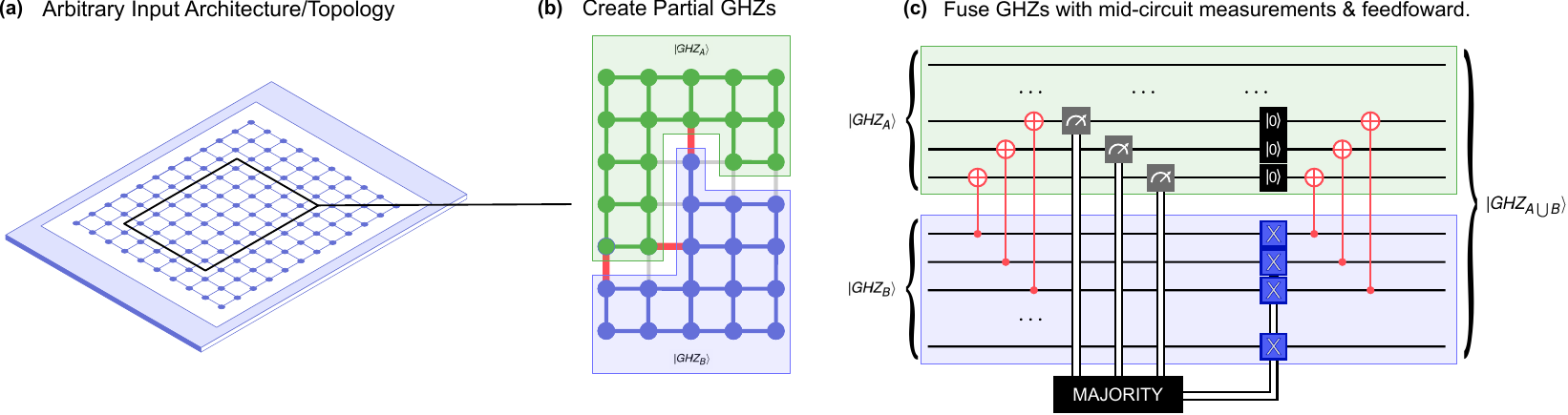}
    \caption{\textbf{Overview of the Group-MV method.}
    \textbf{(a)}~A square lattice quantum processor (illustrated using a grid topology for reference), with a highlighted region indicating the partitioned area.
    \textbf{(b)}~Two adjacent groups, $G_A$ (green) and $G_B$ (blue), each prepared in a local GHZ state. Red edges indicate the boundary links used for majority-vote correction.
    \textbf{(c)}~Dynamic circuit for fusing $\ket{\mathrm{GHZ}_A}$ and $\ket{\mathrm{GHZ}_B}$ into $\ket{\mathrm{GHZ}_{A\cup B}}$.
    CNOTs (red) entangle boundary qubits across groups.
    Mid-circuit measurements on link qubits in $G_B$ feed a classical majority-vote block, which conditionally applies $X$ gates to non-measured qubits in $G_B$.
    Measured qubits are reset to $\ket{0}$ and re-entangled via CNOTs from their partners in $G_A$, restoring full participation in the merged GHZ state.}
    \label{fig:method_overview}
\end{figure*}

\section{Background \& Related Work}
\label{sec:background}
Greenberger-Horne-Zeilinger (GHZ) states~\cite{kafatos_going_1989} are maximally entangled $N$-qubit states of the form
\begin{equation}
\label{eq:ghz_n}\ket{\textrm{GHZ}_N} = \frac{1}{\sqrt{2}}\left( \ket{0}^{\otimes N} + \ket{1}^{\otimes N} \right)
\end{equation}
These states are foundational resources for quantum metrology~\cite{wei_verifying_2020}, sensing~\cite{zang_enhancing_2025}, and device benchmarking~\cite{mooney_generation_2021,kam_characterization_2024}.
However, preparing GHZ states at scale presents a fundamental challenge: circuit depth grows with qubit count, and noise accumulates throughout execution.
Standard approaches such as logarithmic-depth CNOT trees~\cite{cruz_efficient_2019} face this decoherence scaling: as $N$ increases, the probability of error during preparation grows, degrading fidelity before the state can be used.

Dynamic circuits offer a compelling alternative by replacing sequences of quantum gates with mid-circuit measurements and classical feedforward, achieving $\mathcal{O}(1)$ quantum depth~\cite{baumer_efficient_2024}.
Yet this approach trades one error source for another: mid-circuit measurement errors propagate through the feedforward logic, corrupting the final state.
Furthermore, the protocol proposed by Bäumer et al. \cite{baumer_efficient_2024}, which we refer to as \textit{Line Dynamic}, requires embedding qubits in a linear topology, limiting applicability to arbitrary coupling maps.
Measurement-based approaches that extract GHZ states from cluster states~\cite{de_jong_extracting_2024} face similar sensitivity to measurement errors.

However, preparing smaller GHZ states in parallel across qubit groups, then fusing them into a single large state can mitigate depth scaling.
This partition-and-fuse paradigm, developed for quantum networks~\cite{ghaderibaneh_generation_2023,chelluri_resource-_2024}, reduces depth from $\mathcal{O}(\log N)$ to $\mathcal{O}(\log K)$ where $K$ is the largest group size.
Network protocols achieve fusion by distributing Bell pairs between nodes and performing joint measurements.

Adapting partition-and-fuse to single-device settings requires addressing hardware topology: coupling maps constrain which qubits can interact, determining feasible group boundaries.
There is no method that combines partition-and-fuse scalability with arbitrary single-device topologies \emph{and} tolerance to mid-circuit measurement errors.
Our Group-MV protocol addresses precisely this gap: it partitions qubits using topology-aware graph bisection, prepares local GHZ states in parallel, and fuses them via redundant boundary measurements.
To suppress measurement errors at fusion points, the protocol applies majority voting—a classical redundancy technique widely employed in fault-tolerant systems to resolve conflicts among redundant signals~\cite{parhami_voting_1994}—ensuring that isolated measurement errors are outvoted by correct outcomes before feedforward corrections propagate through the circuit.

\section{Proposed Method}
\label{sec:method}

\subsection{Method Overview}
Our approach prepares an $N$-qubit GHZ state by partitioning the qubits into two groups of target size $K$, preparing local GHZ states within each group, and then fusing them via redundant boundary measurements as shown in the example Figure~\ref{fig:method_overview}.
Panel~(a) shows a square lattice coupling graph with a highlighted region indicating the selected qubits, partitioned into adjacent groups. Panel~(b) depicts two such groups, $G_A$ (green) and $G_B$ (blue), each holding a local GHZ state; red edges mark the $L=3$ boundary links that connect them. Panel~(c) shows the dynamic circuit that fuses these local states: CNOTs entangle boundary qubits across groups, mid-circuit measurements feed a majority-vote block, and classically conditioned $X$ gates correct the remaining qubits in $G_B$. Finally, measured qubits are reset and re-entangled to restore full participation in the merged GHZ state.

The key insight behind majority voting is error tolerance: a single mid-circuit measurement error propagates through the feedforward correction. By allocating $L \geq 3$ redundant boundary links between groups and taking the majority vote of their measurement outcomes, we can tolerate up to $\lfloor (L-1)/2 \rfloor$ measurement errors per boundary without affecting the correction decision. This also means $2$ is not an allowed value of $L$.  This robustness comes with tradeoffs. Larger groups ($K$) increase the number of boundaries but also increase local GHZ preparation depth and gate errors. More boundary links ($L$) improve measurement error tolerance but consume additional qubits and introduce extra two-qubit gates. 

\subsection{Algorithm}

Let $G=(V,E)$ be the hardware coupling graph, where vertices represent physical qubits and edges represent native couplers. We select $N$ qubits and partition them into $m=\lceil N/K\rceil$ groups of target size~$K$. An edge $(u,v)\in E$ crossing two groups is a \emph{boundary edge}; one selected to hold a CNOT between two groups is a \emph{boundary link}. We write $L_{\mathrm{eff}}$ for the number of links allocated to a given group pair, with $L\ge1$ the user-specified target redundancy. When $L_{\mathrm{eff}}\ge3$, the correction decision is taken by majority vote over the $L_{\mathrm{eff}}$ mid-circuit measurement outcomes. The majority function returns~1 whenever at least half of its inputs are~1: $\operatorname{MAJ}(c_1,\dots,c_{L_{\mathrm{eff}}}) = \bigvee_{|S|=t}\bigwedge_{i\in S}c_i$, with threshold $t=\lfloor(L_{\mathrm{eff}}+1)/2\rfloor$.

Algorithm~\ref{alg:partition} details the partitioning procedure: qubit selection starts from the center and proceeds via breadth-first search (BFS), then the subgraph is recursively bisected using Kernighan--Lin (KL)~\cite{kernighan_efficient_1970} to produce balanced, connected groups. Algorithm~\ref{alg:circuit} describes the circuit synthesis: local GHZ states are prepared via CNOT trees of $\mathcal{O}(\log K)$ depth, boundary links are distributed via a spanning tree, and cumulative parity tracking generalizes the XOR-chain correction proposed by \cite{baumer_efficient_2024} to the BFS tree over groups. Both algorithms are randomized; we retain the minimum-depth circuit satisfying $L_{\mathrm{eff}}\ge L$ on every boundary.

\subsection{Figure of Merit: Entanglement Witness}

We quantify GHZ state quality via the two-setting entanglement witness inspired by \cite{toth_detecting_2005}:
\begin{equation}
\label{eq:witness}
\mathcal{W}_N(\rho) = \frac{1}{2}\left( P_{|0\rangle^{\otimes N}} + P_{|1\rangle^{\otimes N}} + \langle X^{\otimes N}\rangle \right).
\end{equation}
$\mathcal{W}_N$ only requires two $Z^{\otimes N}$ and $X^{\otimes N}$ measurements and is thus useful to keep simulation time reasonable for large states. Values of $\mathcal{W}_N > \frac{1}{2}$ certify genuine $N$-partite entanglement.
\begin{algorithm}[t]
\caption{Coupling-Graph Partitioning}\label{alg:partition}
\begin{algorithmic}[1]
\Require Coupling graph $G=(V,E)$; qubit count $N$; group size $K$
\Ensure  $m=\lceil N/K\rceil$ connected groups
\State Select $N$ qubits via BFS from the center of $G$
\State Recursively bisect the largest group until $m$ groups
\end{algorithmic}
\end{algorithm}

\begin{algorithm}[t]
\caption{Group-MV GHZ Circuit Synthesis}\label{alg:circuit}
\begin{algorithmic}[1]
\Require Partition $\{G_1,\dots,G_m\}$; boundary pairs $B$; redundancy $L$
\Ensure  Dynamic circuit preparing $N$-qubit GHZ state
\State Prepare local GHZ in each group via parallel CNOT tree \Comment{$\mathcal{O}(\log K)$}
\State Select boundaries via spanning tree; allocate $L_{\textrm{eff}}$ redundant links per boundary
\State Fuse groups: boundary CNOTs $\to$ measure $\to$ majority-vote correction $\to$ re-entangle
\end{algorithmic}
\end{algorithm}


\begin{figure*}[t]
    \centering
    \includegraphics[width=0.86\linewidth]{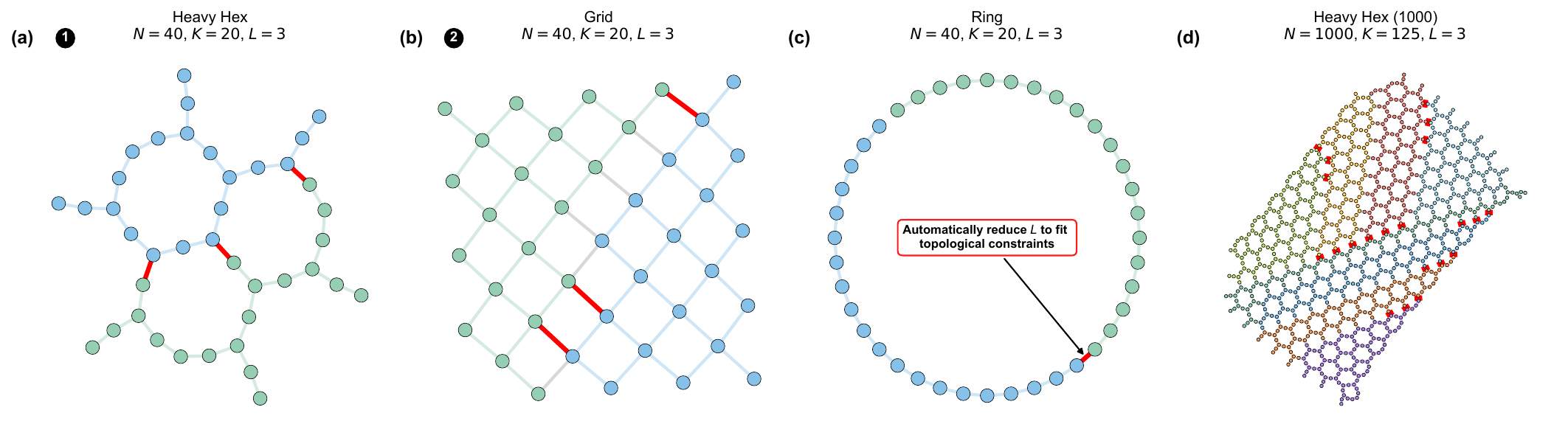}\\
    \vspace{0.3cm}
    \includegraphics[width=0.86\linewidth]{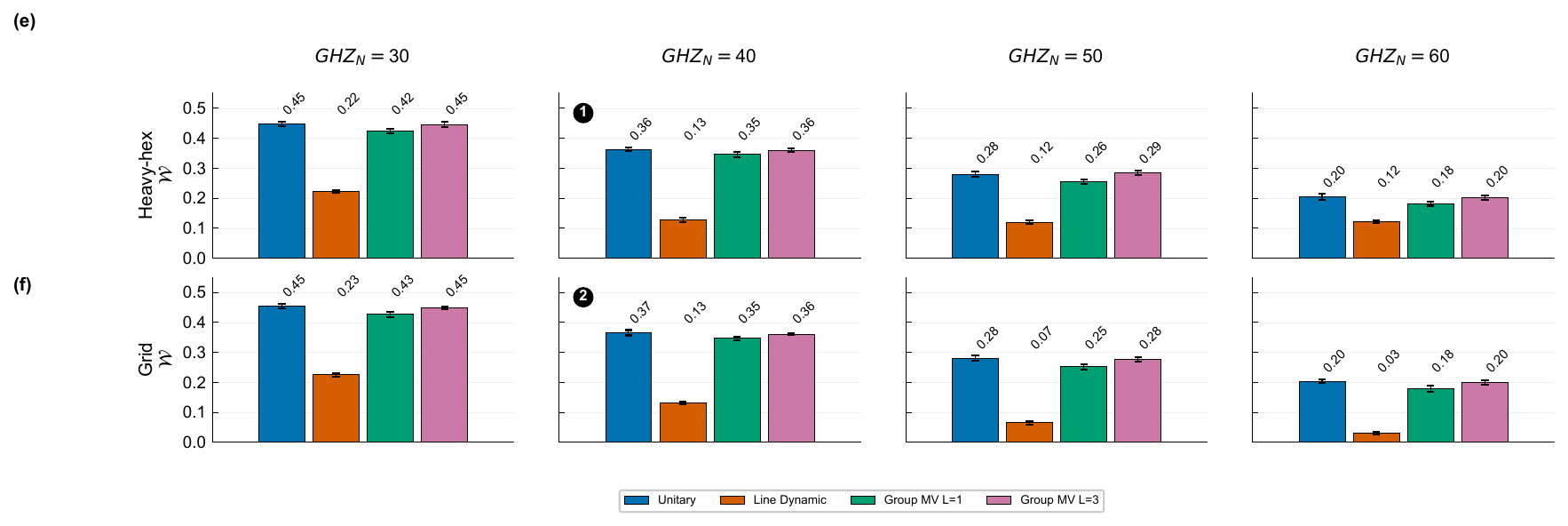}
    
    \caption{\textbf{Graph partitioning and entanglement scaling across topologies.}
\textbf{(a--c)}~Example partitions for Heavy-hex, Grid, and Ring coupling graphs with $N{=}40$ qubits, group size $K{=}20$, and requested boundary redundancy $L{=}3$. Green and blue nodes indicate alternating groups; red edges mark boundary links used for majority-vote fusion. Circled numbers (\circnum{1}, \circnum{2}) indicate correspondence with the $N{=}40$ results in~\textbf{(e)} and~\textbf{(f)}, respectively. In~\textbf{(c)}, the Ring topology cannot support $L{=}3$ due to limited connectivity; our algorithm gracefully degrades to the highest $L$ supported by the architecture.
\textbf{(d)}~Large-scale Heavy Hex example with $N{=}1000$ qubits, $K{=}125$, and $L{=}3$, demonstrating scalability to larger system sizes. For this higher-scale example, the algorithm leverages variable group sizes to support scaling across very large topologies. This partition-and-fuse structure also maps naturally to chiplet-based and heterogeneous quantum architectures, where dense local regions are connected by sparse inter-chip links.
\textbf{(e--f)}~Estimated GHZ entanglement witness $\mathcal{W}_N$ for varying qubit counts on Heavy-hex~\textbf{(e)} and Grid~\textbf{(f)} topologies. Unitary is shown in blue. Line Dynamic \cite{baumer_efficient_2024} (orange) degrades rapidly with increasing $N$. Group-MV with $L{=}1$ (green) and $L{=}3$ (pink) demonstrate improved scaling, with higher redundancy ($L{=}3$) closely tracking the unitary.}
    \label{fig:fidelity_scaling}
\end{figure*}

\section{Results \& Discussion}
\label{sec:results}

\subsection{Experimental Setup}

All algorithms are implemented in Python using the Qiskit framework; source code is available at \url{https://github.com/jbwaring/group-mv}.
We evaluate Group-MV using Qiskit's AerSimulator with the stabilizer simulation method.
The noise model supposes isotropic depolarizing errors over the whole hardware graph: $p_\text{CX} = 0.01\%$ for two-qubit gates and $p_\text{1Q} = 0.01\%$ for single-qubit gates.
Readout errors with $p_\text{RO} = 5\%$ probability of a bit-flip on a measurement are considered.
M3 readout error mitigation~\cite{nation_scalable_2021} is enabled throughout. Overall, the use of M3, as well as setting intentionally low gate errors, is designed to isolate the effects of readout errors on the mid-circuit measurements alone.

We test our method on two topologies: (i)~\emph{Heavy-hex} and (ii)~\emph{Grid} (square lattice).
Qubit counts range is $N \in \{30, 40, 50, 60\}$ with target group size $K{=}20$. We compare the values of $L$: 1 and 3 to evaluate the gains from majority-voting.
Each configuration is repeated 10 times, and $\mathcal{W}_N$ is computed for each step. Figure~\ref{fig:fidelity_scaling} \textbf{(c)} shows how our algorithm gracefully handles a sparse coupling graph by returning a lower $L$ than requested.

\subsection{Entanglement Results}

Figure~\ref{fig:fidelity_scaling} compares GHZ entanglement witness $\mathcal{W}_N$ across methods and topologies.
The \emph{Unitary} baseline provides a benchmark limited only by noise accumulation.
\emph{Line Dynamic} implements Bäumer et al.'s 1D chain protocol~\cite{baumer_efficient_2024}. Group-MV with $L{=}3$ boundary links closely tracks the unitary baseline across all tested $N$ values.
At $N{=}30$, Group-MV L=3 achieves $\mathcal{W}_N = 0.45$ (Heavy-hex, Fig. \ref{fig:fidelity_scaling} \textbf{(e)}~) and 0.45 (Grid, Fig. 2 \textbf{(f)}~), matching the unitary references. For $N=40$, we show a correspondence between Heavy-hex partitioning for $L=3$ and its corresponding $\mathcal{W}_N$ identified by \circnum{1}. Likewise, \circnum{2} presents the same for a grid topology.
At larger scales, Group-MV systematically outperforms Line Dynamic. At $N{=}60$, Group-MV L=3 yields 0.20 (Heavy-hex) and 0.20 (Grid), matching unitary values.

In contrast, Line Dynamic degrades faster with $N$ than both Group-MV and Unitary methods, particularly on Grid topologies. This is largely due to the higher number of non-mitigated measurements it requires. Group-MV also exhibits consistent behavior across both Heavy-hex and Grid architectures, with differences typically below 5\% at matched $N$ values.
Unlike methods that impose fixed circuit structures, Group-MV leverages whatever local connectivity exists within each partition, ensuring its architectural independence.

\subsection{Majority Voting Efficacy}

For $N \geq 30$, we compare $L{=}1$ (single boundary) against $L{=}3$ (majority-voted triple boundary).
The $L{=}3$ configuration consistently outperforms $L{=}1$: at $N{=}50$ on Heavy-hex, $\mathcal{W}_N$ improves from 0.26 ($L{=}1$) to 0.29 ($L{=}3$).
At $N{=}60$ on Grid, the improvement is 0.18 to 0.20.

This gap suggests that majority voting does effectively mitigates mid-circuit measurement errors that would otherwise propagate through the classical feedforward path.
Mid-circuit measurement errors are particularly damaging in dynamic circuits because incorrect classical bits propagate through feedforward operations to corrupt all subsequent qubits.
By taking a majority vote over three independent measurements, Group-MV tolerates single-bit flip errors at the fusion boundary. The improvement from $L{=}1$ to $L{=}3$ is modest (typically 5--15\%) because the dominant error source remains gate noise accumulated during local GHZ preparation, not measurement errors at fusion points.
Nevertheless, as readout error rates on hardware often exceed gate error rates, the majority-vote mechanism provides a meaningful robustness improvement.

\subsection{Fidelity Validation}

To corroborate the witness-based results, we estimate state fidelity via stabilizer sampling on Heavy-hex with $N{=}30$.
The fidelity estimates are: Unitary $F = 0.093 \pm 0.016$, Line Dynamic $F = 0.049 \pm 0.007$, and Group-MV $L{=}3$ $F = 0.119 \pm 0.014$.
Group-MV achieves the highest fidelity, outperforming both the unitary baseline and Line Dynamic, consistent with the witness-based comparisons.
\subsection{Opportunities \& Challenges}
While our results demonstrate measurable gains for Group-MV at modest scales ($N < 100$), we anticipate more pronounced advantages at larger qubit counts or on non-monolithic architectures. 
The parallel preparation strategy limits local depth to $\mathcal{O}(\log K)$ regardless of total system size and our algorithm is well capable of handling topologies of thousands of qubit as shown Figure 2 \textbf{(d)}. Sufficiently large groups could also enable group boundaries supporting larger $L$ values, such as 5.
However, simulation scales poorly, preventing systematic design-space exploration.
In particular, quantifying the overhead of mid-circuit measurements---including reset latency and conditional gate delays---relative to the depth savings they enable remains an open question that simulation alone cannot resolve.
On current hardware, mid-circuit operations incur latencies that may offset depth reductions; understanding this tradeoff requires empirical characterization on actual devices or the development of specialized simulators that model measurement-induced delays faithfully. These limitations points to a clear path forward: hardware validation on superconducting processors will not only verify our simulation results but also enable exploration of the $(K, L)$ design space under realistic timing constraints.
\vspace*{14pt}
\section{Conclusion}
\label{sec:conclusion}
We have demonstrated an approach for preparing large-scale GHZ states that adapts
automatically to arbitrary hardware topologies without requiring manual circuit design.
By partitioning the coupling graph and fusing local GHZ states via majority-voted
boundary measurements, we can mitigate the measurement errors that would otherwise
corrupt classical feedforward, maintaining fidelity within 3\% of unitary preparation
even at 60 qubits. The partition-and-fuse structure scales naturally to systems
exceeding 1000 qubits through variable group sizing.
Beyond monolithic devices, Group-MV naturally maps onto chiplet-based and heterogeneous quantum architectures,
where dense local processor regions are connected by sparse inter-chip links.
\clearpage
\bibliographystyle{IEEEtran}
{\scriptsize
\bibliography{references}
}

\end{document}